\pdfoutput=1
\documentclass[twocolumn,superscriptaddress,aps,preprintnumbers,amsmath,amssymb,prd,nofootinbib,floatfix]{revtex4-2}

\usepackage{amsmath}
\usepackage{amssymb}
\usepackage{amsfonts}
\usepackage{mathrsfs}
\usepackage{graphicx}
\usepackage{xcolor}
\usepackage{physics}
\usepackage{fourier}
\usepackage{hyperref}
\usepackage{bm}
\usepackage{booktabs}

\definecolor{rossoferrari}{HTML}{D9073D}
\definecolor{lightblue}{rgb}{0.284602,0.317763,0.963947}
\hypersetup{
  setpagesize=false,
  bookmarksnumbered=true,
  bookmarksopen=true,
  colorlinks=true,
  linkcolor=lightblue,
  urlcolor=rossoferrari,
  citecolor=rossoferrari,
  linktocpage=false,
  pdftitle={Tunneling from a Slowly Rolling Multifield Background Using Tadpole Subtraction},
  pdfauthor={Masaki Yamada}
}

\newcommand{\ii}{\mathrm{i}}
\newcommand{\mpl}{M_{\mathrm{Pl}}}
\newcommand{\ts}{\mathrm{TS}}
\newcommand{\TS}{\mathrm{TS}}
\newcommand{\eff}{\mathrm{eff}}
\newcommand{\bphi}{\bar\phi}

\newcommand{\inner}[2]{\left\langle #1,#2\right\rangle}

\begin{document}

\title{Tunneling from a Slowly Rolling Multifield Background Using Tadpole Subtraction}

\author{Masaki Yamada}
\email{m.yamada@tohoku.ac.jp}
\affiliation{Department of Physics, Tohoku University, Sendai, Miyagi 980-8578, Japan}

\preprint{TU-1317}

\date{\today}

%%%%%%%%%%%%%%%%%%%%%%%%%%%%%%%%%%%%%%%%%%%%%%%%%%%%%%%%%%%%%%%%%%%%%%%%%%%%%%%%%%%%%%%%%%%%%%%%%%%%

\begin{abstract}
We study semiclassical tunneling in scalar field theories with slowly evolving homogeneous backgrounds.
We formulate the problem by combining tadpole subtraction with an adiabatic expansion: the former maps the rolling system to an instantaneous multi-field tunneling problem, while the latter systematically tracks the slow evolution of the background.
This formulation is particularly useful for tunneling during slow-roll inflation, where the tunneling direction may be transverse to the rolling inflaton direction.  
For a time-reflection-symmetric bounce on a fixed on-shell geometry, we show that the real tunneling exponent receives no direct correction linear in the rolling velocity.  The tunneling rate nevertheless drifts along the rolling trajectory because the frozen tadpole-subtracted potential evolves. 
We also find that the exponent approaches a finite but nonanalytic limit as the tangential curvature vanishes, suggesting that a weakly tachyonic tangential direction can yield an exponent of the same order.
\end{abstract}

\maketitle

%%%%%%%%%%%%%%%%%%%%%%%%%%%%%%%%%%%%%%%%%%%%%%%%%%%%%%%%%%%%%%%%%%%%%%%%%%%%%%%%%%%%%%%%%%%%%%%%%%%%
\section{Introduction}
\label{sec:introduction}
%%%%%%%%%%%%%%%%%%%%%%%%%%%%%%%%%%%%%%%%%%%%%%%%%%%%%%%%%%%%%%%%%%%%%%%%%%%%%%%%%%%%%%%%%%%%%%%%%%%%

False-vacuum decay is a universal mechanism of metastability in systems with many degrees of freedom, with applications ranging from thermal and quantum nucleation in condensed-matter systems to phase transitions in particle physics and cosmology~\cite{Langer:1969bc,Caldeira:1981rx,Zenesini:2023afv,Vodeb:2024tvo}. In relativistic field theory, the standard theory assumes a stationary metastable state and computes the leading semiclassical exponent from the Euclidean bounce~\cite{Coleman:1977py,Callan:1977pt}. This method has been extended to multi-field tunneling~\cite{Konstandin:2006nd,Blum:2016ipp,Wainwright:2011kj,Masoumi:2016wot,Espinosa:2018szu,Guada:2020xnz,Espinosa:2023oml}, finite-temperature transitions~\cite{Linde:1980tt,Linde:1981zj}, and gravitational effects~\cite{Coleman:1980aw}, and forms the basis for most studies of cosmological first-order phase transitions.

There has recently been growing interest in vacuum decay beyond the stationary false-vacuum setup, where either the initial state or the effective tunneling potential is time dependent.  These problems have been studied using time-dependent bounce methods, complex-time WKB techniques, real-time and functional approaches, Lorentzian path-integral methods, and finite-time or slowly varying background expansions~\cite{Widrow:1991xu,Keski-Vakkuri:1996lbi,Simon:2009nb,Bramberger:2016yog,Darme:2017wvu,Hertzberg:2019wgx,Ai:2019fri,Darme:2019ubo,Hayashi:2021kro,Draper:2023fkz,Steingasser:2023gde,Batini:2023zpi,Steingasser:2024ikl,Garbrecht:2024end,Lin:2025wgc,Janssen:2026qwq,Yuwen:2026nsb}.

Of particular interest is the adiabatic regime, where the background evolves slowly compared with the characteristic time scale of the tunneling event. Phase transitions during inflation provide an important example of such a vacuum-transition problem.%
\footnote{
Throughout this work we focus on localized, sub-horizon tunneling events. This should be distinguished from transitions described by stochastic inflation~\cite{Starobinsky:1994bd} or by Hawking--Moss-type saddle points~\cite{Hawking:1981fz,Weinberg:2006pc}, where long-wavelength light fields fluctuate over Hubble-scale patches.
}
Vacuum bubbles nucleated during inflation can seed primordial black holes or baby-universe configurations~\cite{Garriga:2015fdk,Deng:2017uwc,Deng:2018cxb,Deng:2020mds,Kleban:2023ugf}, and first-order transitions in spectator sectors may generate characteristic gravitational-wave signals~\cite{Adams:1990ds,Ashoorioon:2015hya,Jiang:2015qor,Wang:2018caj,An:2020fff,An:2022cce,Tong:2023krn,An:2023jxf,Sugeno:2025kwx,Hu:2025xdt,Zou:2026wzi,Ghoderao:2026lvz}, CMB temperature-anisotropy signatures~\cite{Firouzjahi:2017ssb,Fujikura:2025xgl,Cembranos:2026cat}, or dark matter~\cite{Cheng:2026npt}. In these scenarios the tunneling field is often transverse to the inflaton direction, while the inflaton controls the slow time evolution of the effective potential. The background is therefore a slow-roll trajectory, not a static false vacuum.%
\footnote{
This should be distinguished from open-inflation scenarios~\cite{Freivogel:2005vv,Sugimura:2011tk}, where the system first tunnels from a static false vacuum and slow-roll inflation subsequently occurs inside the bubble.
}
Nevertheless, most phenomenological studies prescribe an instantaneous transverse bounce at a fixed value of the inflaton. The validity and limitations of this prescription remain to be clarified.

In this Letter we propose a formulation of tunneling from a rolling background based on what we call the \emph{tadpole subtraction method}. We decompose the scalar fields into a time-dependent homogeneous background and fluctuations around it, and subtract the tadpole associated with the rolling background. The fluctuation fields then have a stationary initial configuration, while the non-stationarity of the original problem appears as explicit time dependence in the effective potential. For a slow-roll background, this induced time dependence is parametrically small. The reformulation therefore allows us to apply a controlled semiclassical expansion around a static initial point, while systematically keeping track of the slow evolution of the background.

%%%%%%%%%%%%%%%%%%%%%%%%%%%%%%%%%%%%%%%%%%%%%%%%%%%%%%%%%%%%%%%%%%%%%%%%%%%%%%%%%%%%%%%%%%%%%%%%%%%%
\section{Tadpole subtraction for a rolling background}
\label{sec:tadpole}
%%%%%%%%%%%%%%%%%%%%%%%%%%%%%%%%%%%%%%%%%%%%%%%%%%%%%%%%%%%%%%%%%%%%%%%%%%%%%%%%%%%%%%%%%%%%%%%%%%%%

\subsection{Tadpole-subtracted action on a fixed geometry}

We consider $N$ canonical scalar fields $\phi^I$ ($I=1,2,\dots,N$) on a fixed spacetime geometry.  The scalar action is
\begin{align}
  S[\phi]
  ={}&\int\dd^4x\sqrt{-g}
  \left[
    -\frac12\delta_{IJ}g^{\mu\nu}
      \partial_\mu\phi^I\partial_\nu\phi^J
    -V(\phi)
  \right].
  \label{eq:einstein_scalar_action}
\end{align}
In the main part of this paper, we take the background geometry to be a spatially flat FLRW spacetime with scale factor $a(t)$ and Hubble parameter $H=\dot a/a$.  The homogeneous background fields $\bphi^I(t)$ satisfy
\begin{align}
  \ddot{\bphi}^I+3H\dot{\bphi}^I+ V^{,I}(\bphi)=0,
  \label{eq:background_eom}
\end{align}
where a dot denotes a derivative with respect to physical time $t$, and $V^{,I} = \delta^{IJ} V_{,J} = \delta^{IJ} \partial V / \partial \bphi^J$.

To isolate the effect of the rolling background, we decompose the fields as
\begin{align}
  \phi^I(t,\bm x)=\bphi^I(t)+\eta^I(t,\bm x).
  \label{eq:field_split}
\end{align}
Subtracting the homogeneous background action, we obtain
\begin{align}
  &S[\bphi+\eta]-S[\bphi]
  \nonumber\\
  &=
  \int \dd^4 x\,a^3
  \bigg[
    \delta_{IJ}\dot{\bphi}^I\dot\eta^J
    +\frac{\dot\eta^2}{2}
    -\frac{\left( \bm\nabla\eta \right)^2}{2a^2} 
    -V(\bphi+\eta)+V(\bphi)
  \bigg]
  \nonumber\\
  &=
  \left[
    \int\dd^3x\,a^3\dot{\bphi}_I\eta^I
  \right]_{t_i}^{t_f}
  +
  \int\dd^4 x\,a^3
  \left[
    \frac{\dot\eta^2}{2}
    -\frac{(\bm\nabla\eta)^2}{2a^2}
    -\Delta U_{\ts}(\eta;t)
  \right],
  \label{eq:exact_subtracted_action}
\end{align}
where we integrated the kinetic cross term by parts and used Eq.~(\ref{eq:background_eom}). 
Here and below, contractions of field-space indices are taken with $\delta_{IJ}$, 
and 
we have defined
\begin{align}
  \Delta U_{\ts}(\eta;t)
  &\equiv U_{\ts}(\eta;t)-U_{\ts}(0;t),
  \\
  U_{\ts}(\eta;t)
  &\equiv
  V(\bphi(t)+\eta)-V_{,I}(\bphi(t))\eta^I .
  \label{eq:tadpole_subtracted_potential}
\end{align}
By construction, $\partial U_{\ts}/\partial\eta^I$ vanishes at $\eta=0$.  Thus the force acting on the rolling background is removed from the fluctuation action, while all higher-order terms of the original potential retain their dependence on the instantaneous background value.  We refer to this procedure as \emph{tadpole subtraction}, in the sense that it removes the term linear in the finite fluctuation.  This should not be confused with tadpole renormalization at the loop level.  Equation~(\ref{eq:exact_subtracted_action}) is exact at arbitrary fluctuation amplitude, provided that the scalar background is on shell and the geometry is held fixed.

The boundary term in Eq.~(\ref{eq:exact_subtracted_action}) contributes only a phase to the Lorentzian transition amplitude and therefore does not affect the real tunneling exponent.  Its physical interpretation, as well as its relation to the generalized escape paths of Ref.~\cite{Darme:2019ubo}, is discussed in App.~\ref{app:GEP}.

\subsection{Frozen exponent at a candidate nucleation time}
\label{sec:frozen}

The tadpole-subtracted potential in Eq.~(\ref{eq:tadpole_subtracted_potential}) depends explicitly on time through the rolling background $\bphi(t)$.  We now assume that the background evolves slowly enough that both the effective potential and the bubble profile change little over the formation time of a tunneling event.  At leading adiabatic order, we therefore freeze the background at a candidate nucleation time $t=t_*$ and solve the resulting time-independent tunneling problem in the standard way.  Corrections associated with the time dependence during a single event will be discussed in Sec.~\ref{sec:time_corrections}.

We define
\begin{align}
  \bphi_*^I \equiv \bphi^I(t_*),
  \qquad
  U_{\rm TS,*}(\eta) \equiv U_{\ts}(\eta;t_*),
  \label{eq:frozen_definitions}
\end{align}
and, without loss of generality, normalize the scale factor such that $a(t_*)=1$.
When the frozen-background problem admits a real, time-reflection-symmetric $O(4)$ bounce $\eta_b^I(\rho)$, the leading exponent is
\begin{align}
  B_0(t_*)
  &=
  S_E[\eta_b;t_*]-S_E[0;t_*]
  \nonumber\\
  &=
  2\pi^2\int_0^\infty\dd\rho\,\rho^3
  \left[
    \frac12\delta_{IJ}
    \eta_b^{I\prime}\eta_b^{J\prime}
    +\Delta U_{\rm TS,*}(\eta_b)
  \right],
  \label{eq:O4_action}
\end{align}
where $\rho$ is the $O(4)$ radial coordinate, 
and the bounce $\eta_b$ satisfies
\begin{align}
  \eta_b^{I\prime\prime}
  +\frac{3}{\rho}\eta_b^{I\prime}
  &=
  \frac{\partial U_{\rm TS,*}}{\partial\eta_b^I},
  \nonumber\\
  \eta_b^{I\prime}(0)&=0,
  \qquad
  \eta_b^I(\infty)=0.
  \label{eq:multifield_bounce_eom}
\end{align}
Here a prime denotes a derivative with respect to $\rho$.%
\footnote{
The homogeneous rolling kinetic energy is encoded in the extracted phase in Eq.~(\ref{eq:scalar_background_phase}) and does not enter the frozen bulk potential.
}

The boundary-value problem is therefore the standard multi-field tunneling problem.  The exponent can be computed using established path-deformation, multiple-shooting, and action-minimization algorithms~\cite{
Konstandin:2006nd,Wainwright:2011kj,Masoumi:2016wot,Espinosa:2018szu,Guada:2020xnz}.
In the numerical example below, we use a finite-difference Newton method.

The instantaneous nucleation-rate density takes the schematic form
\begin{align}
  \Gamma(t_*)\simeq\mathcal A(t_*)\exp[-B(t_*)],
  \label{eq:instantaneous_rate}
\end{align}
where the prefactor $\mathcal A$ is not evaluated in this paper.  Even in the frozen approximation, the exponent varies from one candidate nucleation time to another such as
\begin{align}
  B(t_*)=B[\bphi(t_*)],
  \qquad
  \dot B
  =\dot{\bphi}_*^I\frac{\partial B}{\partial\bphi_*^I}.
  \label{eq:interevent_drift}
\end{align}
This order-$\dot{\bphi}_*$ variation represents an inter-event drift of the instantaneous rate.  It should not be confused with a correction generated by the time evolution during a single bounce.

\subsection{Time dependence during one tunneling event}
\label{sec:time_corrections}

We next isolate the correction due to the explicit time dependence during a single tunneling event.  Expanding the tadpole-subtracted potential around $t=t_*$ gives
\begin{align}
  \Delta U_{\ts}(\eta;t)
  &=
  \Delta U_{\rm TS,*}(\eta)
  +(t-t_*)\Delta U_1(\eta)+\cdots,
  \label{eq:potential_time_expansion}
\end{align}
with
\begin{align}
  \Delta U_1(\eta)
  &=
  \dot{\bphi}_*^I
    \frac{\partial U_{\ts}}{\partial\bphi^I}(\eta;t_*)
  -
  \dot{\bphi}_*^I
    \frac{\partial U_{\ts}}{\partial\bphi^I}(0;t_*)
  \nonumber\\
  &=
  \dot{\bphi}_*^I
  \left[
    V_{,I}(\bphi_*+\eta)
    -V_{,I}(\bphi_*)
    -V_{,IJ}(\bphi_*)\eta^J
  \right].
  \label{eq:U1_definition}
\end{align}
The second term in Eq.~(\ref{eq:potential_time_expansion}) gives the first-order adiabatic correction to the tunneling problem.

We evaluate this correction by the Wick rotation around $t=t_*$ along the vertical contour, $t=t_*-\ii\tau$.  The explicit first-order perturbation of the Euclidean action, evaluated on the zeroth-order bounce $\eta_b$, is
\begin{align}
  \delta S_E^{(1)}
  &=
  -\ii\int\dd^4x_E\,\tau\,
  \Delta U_1\bigl(\eta_b(\rho)\bigr)
  \nonumber\\
  &=0,
  \label{eq:linear_event_correction_zero}
\end{align}
because the integrand is odd in $\tau$.  Thus the first-order adiabatic correction is purely imaginary, or vanishes after the symmetric Euclidean integration, and does not contribute to the real tunneling exponent for a real, time-reflection-symmetric bounce.

Let $R_b$ denote the characteristic bubble size.  The leading corrections to the scalar tunneling exponent are then proportional to
$\ddot{\bphi}_*R_b^2$ and
$\dot{\bphi}_*^2R_b^2$.
These second-order effects are negligible in the slow-roll regime considered here.

%%%%%%%%%%%%%%%%%%%%%%%%%%%%%%%%%%%%%%%%%%%%%%%%%%%%%%%%%%%%%%%%%%%%%%%%%%%%%%%%%%%%%%%%%%%%%%%%%%%%
\section{A two-field slow-roll example}
\label{sec:explicit_model}
%%%%%%%%%%%%%%%%%%%%%%%%%%%%%%%%%%%%%%%%%%%%%%%%%%%%%%%%%%%%%%%%%%%%%%%%%%%%%%%%%%%%%%%%%%%%%%%%%%%%

\subsection{Model and frozen tadpole-subtracted potential}

We now apply the construction to a two-field model with a straight slow-roll trajectory and a barrier in the transverse direction.  We denote the field-space components by $\phi^1\equiv\phi$ and $\phi^2\equiv\psi$.  The field $\phi$ is the rolling, tangential field, while $\psi$ is the transverse tunneling field.  We choose the classical rolling trajectory to lie at $\psi=0$.

Since we are interested in slow-roll inflation, we use a local expansion of the inflaton potential around a reference field value, which we set to $\phi=0$.  We consider
\begin{align}
  V(\phi,\psi)
  ={}&V_0-\epsilon\phi
  +\frac12\mu_\phi^2\phi^2
  +m^2\psi^2\left(1+\frac{\phi}{v_1}\right)
  \nonumber\\
  &-g_3m\psi^3\left(1+\frac{\phi}{v_2}\right)
  +\lambda\psi^4\left(1+\frac{\phi}{v_3}\right).
  \label{eq:explicit_potential}
\end{align}
The mass dimensions are $[m]=[v_i]=[\mu_\phi]=1$, $[\epsilon]=3$, and $[g_3]=[\lambda]=0$.  We first consider $\mu_\phi^2\geq0$ and discuss a weakly tachyonic continuation separately below.

At a candidate nucleation time $t_*$, we define $\phi_*\equiv\bar\phi(t_*)$ and write
\begin{align}
  \phi=\phi_*+\eta .
\end{align}
Using $V_{,\phi}(\phi_*,0)=-\epsilon+\mu_\phi^2\phi_*$ and $V_{,\psi}(\phi_*,0)=0$,
the frozen tadpole-subtracted potential is obtained from Eq.~(\ref{eq:tadpole_subtracted_potential}) such as 
\begin{align}
  \Delta U_{\rm TS,*}(\eta,\psi)
  ={}&
  \frac12\mu_\phi^2\eta^2
  +m^2\psi^2\left(1+\frac{\phi_*+\eta}{v_1}\right)
  \nonumber\\
  &-g_3m\psi^3\left(1+\frac{\phi_*+\eta}{v_2}\right)
  +\lambda\psi^4\left(1+\frac{\phi_*+\eta}{v_3}\right).
  \label{eq:explicit_delta_U_TS}
\end{align}

\subsection{Numerical results}
\label{sec:numerical_method}

We solve the coupled bounce equations numerically with the boundary conditions in Eq.~(\ref{eq:multifield_bounce_eom}).  The radial equations are discretized on a uniform grid $\rho_i=ih$ ($i=0,1,\dots,N$), with $R=Nh$, using second-order centered differences.  At the origin, regularity is imposed by
\begin{align}
  \left(q''+\frac{3}{\rho}q'\right)_{\rho=0}
  =
  \frac{8[q(h)-q(0)]}{h^2}
  +\order{h^2},
  \label{eq:num-origin}
\end{align}
where $q\in\{\eta,\psi\}$.  We solve the nonlinear system by damped Newton iteration, using a $2\times2$ block-tridiagonal solver at each linear step, and stop the iteration when the largest absolute residual is below $2\times10^{-10}$.

The asymptotic solution for the tangential field is proportional to $K_1(\mu_\phi\rho)/\rho$, 
where $K_i(x)$ is the modified Bessel function of the second kind. 
This motivates the Robin boundary condition
\begin{align}
  \eta'(R)+\kappa(\mu_\phi,R)\eta(R)&=0,
  \nonumber\\
  \kappa(\mu_\phi,R)
  &=
  \mu_\phi
  \frac{K_0(\mu_\phi R)}{K_1(\mu_\phi R)}
  +\frac{2}{R},
  \label{eq:num-robin}
\end{align}
while we impose $\psi(R)=0$.  
The contribution to the exponent from the exterior region is given by 
\begin{align}
  B_{\mathrm{tail}}
  \simeq 2 \pi^2 \int_R^\infty \dd \rho\, \rho^3 \left[ 
  \frac12\eta'^2+\frac12 \mu_\phi^2 \eta^2
    \right]
 = 
  \pi^2R^3\kappa(\mu_\phi,R)\eta(R)^2 .
  \label{eq:num-tail-action}
\end{align}
The full exponent is then
\begin{align}
  B_{\rm full}
  =
  2\pi^2\int_0^R \dd\rho\,\rho^3
  \left[
    \frac12\eta'^2+\frac12\psi'^2
    +\Delta U_{\rm TS,*}(\eta,\psi)
  \right]
  +B_{\mathrm{tail}},
  \label{eq:num-full-action}
\end{align}
where the finite-interval integral is evaluated on the radial grid.

To illustrate the effect of tangential relaxation, we compare the fully coupled bounce with two approximations.  First, setting $\eta=0$ gives a constrained transverse bounce, which is a single-field tunneling problem and can be solved by shooting.  We denote its solution and exponent by $\psi_0(\rho)$ and $B_0$, respectively.  Second, for weak tangential coupling, we compute the leading tangential response $\eta_1(\rho)$ around the constrained profile $\psi_0(\rho)$.  It satisfies
\begin{align}
  \eta_1''
  +\frac{3}{\rho}\eta_1'
  -\mu_\phi^2\eta_1
  &=
  J(\psi_0),
  \label{eq:model_linear_response_eom}
  \\
  J(\psi)
  &=
  \frac{m^2}{v_1}\psi^2
  -\frac{g_3m}{v_2}\psi^3
  +\frac{\lambda}{v_3}\psi^4,
  \label{eq:model_J}
\end{align}
with $\eta_1'(0)=0$ and $\eta_1(\infty)=0$.  The corresponding linear-response contribution to the exponent is
\begin{align}
  B_1
  &=
  2\pi^2\int_0^\infty\dd\rho\,\rho^3
  \left[
    \frac12\eta_1'^2
    +\frac12\mu_\phi^2\eta_1^2
    +\eta_1J(\psi_0)
  \right]
  \nonumber\\
  &=
  -\pi^2\int_0^\infty\dd\rho\,\rho^3
  \left(\eta_1'^2+\mu_\phi^2\eta_1^2\right),
  \label{eq:model_linear_response_action}
\end{align}
where the second equality follows from the linearized equation of motion.

For the numerical example, we measure all dimensionful quantities in units of $m$ and set
\begin{align}
  v_1=v_2=v_3=1,
  \qquad
  g_3=3,
  \qquad
  \lambda=1 .
  \label{eq:num-parameters}
\end{align}
We use a box size $R=60$ and grid spacing $h=0.01$.

\begin{figure}[!t]
  \centering
  \includegraphics[width=0.48\textwidth]{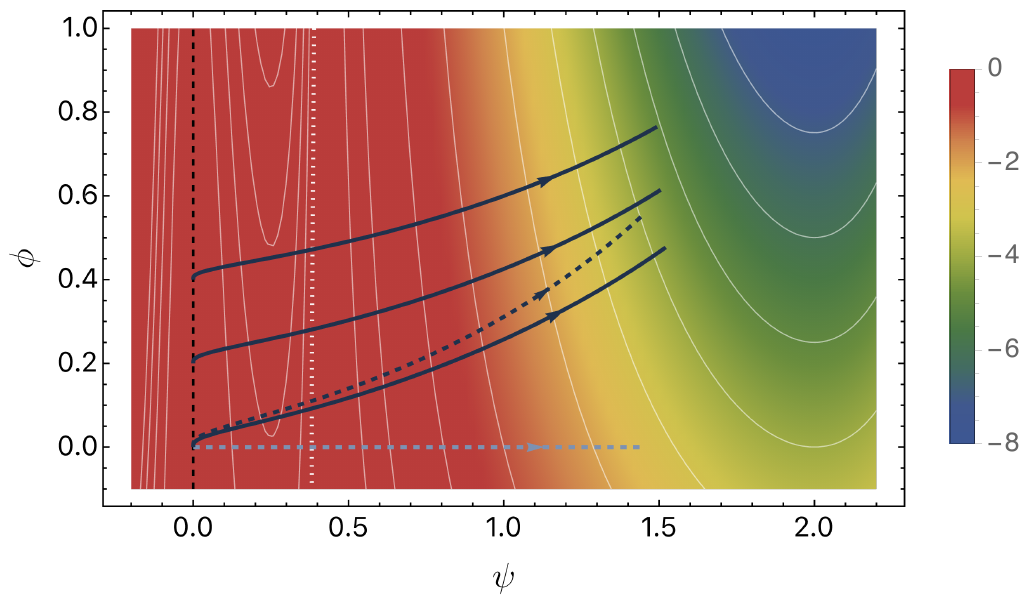}
  \caption{
Field-space trajectories for $\mu_\phi^2=10^{-2}$ overlaid on the frozen tadpole-subtracted potential $\Delta U_{\rm TS,*}(\eta,\psi)$.
The vertical black dashed line denotes the classical inflaton trajectory, $\psi=0$.
The horizontal and diagonal dashed curves show the fixed-field and linear-response approximations for $\phi_* = 0$, respectively.
The solid curves show the fully coupled relaxation trajectories for $\phi_*=0$, $0.2$, and $0.4$, ordered from bottom to top.
The color bar indicates $\Delta U_{\rm TS,*}$ for $\phi_*=0$.
The white solid curves denote the contours $\Delta U_{\rm TS,*}=-7,-6,-5,-4,-3,-2,-1,-0.5,-0.2,-0.1,0.01,0.02,0.03,0.04$.
The white dotted curve denotes $\Delta U_{\rm TS,*}=0$.
}
  \label{fig:fieldvalue}
\end{figure}

The coupled solution for $\mu_\phi^2=10^{-2}$ is shown by the solid curves in Fig.~\ref{fig:fieldvalue} for $\phi_* = 0, 0.2$, and $0.4$.  
The horizontal and diagonal dashed curves show the constrained transverse bounce and the one including the leading tangential response for $\phi_* = 0$, respectively.
The corresponding exponents are
\begin{align}
  B_0&=12.5,
  \\
  B_0+B_1&=10.3,
  \\
  B_{\mathrm{full}}&=10.8.
  \label{eq:num-actions-d001}
\end{align}
Thus the fully coupled bounce lowers the constrained transverse action by about $13\%$.  The linear-response approximation captures the sign of the effect but overestimates the reduction.

\subsection{Dependence on the tangential curvature}

\begin{figure}[!htbp]
  \centering
  \includegraphics[width=0.92 \linewidth]{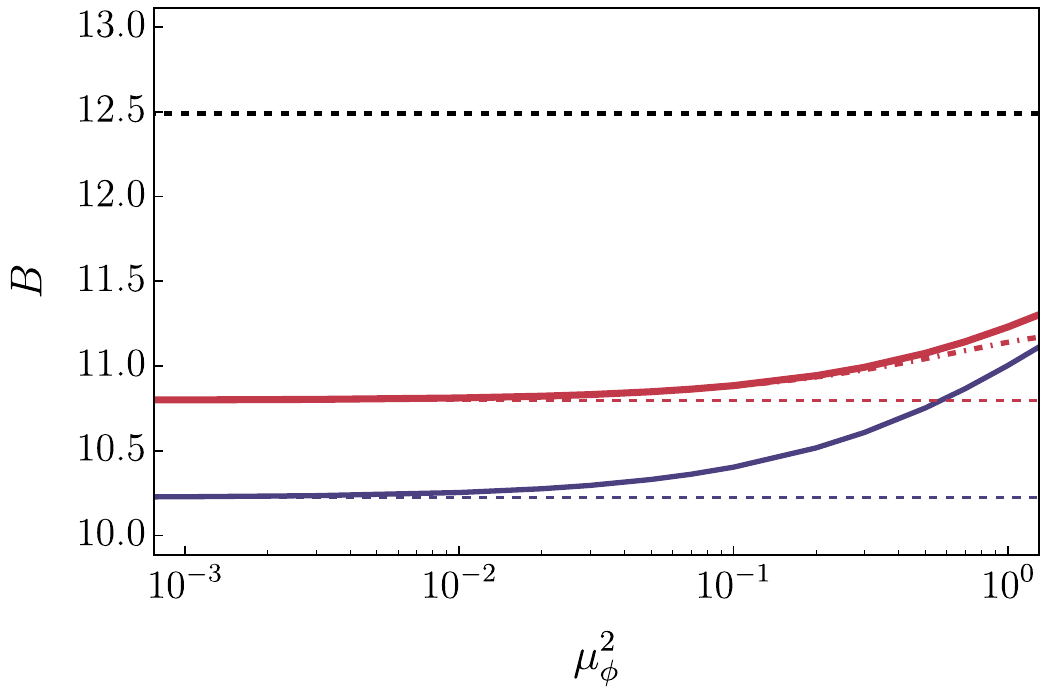}
  \caption{
Approach to the massless tangential limit from positive $\mu_\phi^2$.
The black dotted horizontal line shows the constrained action $B_0$, which is independent of $\mu_\phi^2$.
The red curve shows the fully coupled action $B_{\rm full}(\mu_\phi^2)$, and the blue curve shows the linear-response estimate $B_0+B_1(\mu_\phi^2)$.
The dashed horizontal lines indicate the corresponding one-sided limits as $\mu_\phi^2\to0^+$.
}
  \label{fig:num-massless-limit}
\end{figure}

Figure~\ref{fig:num-massless-limit} shows the dependence of the exponent on positive $\mu_\phi^2$ for $\phi_* = 0$.  The exponent approaches a finite constant as $\mu_\phi^2\to0^+$.  To see this asymptotic form, define the massless source charge
\begin{align}
  Q_{\mathrm{full}}
  =
  2\pi^2\int_0^\infty\dd\rho\,\rho^3J(\psi),
  \label{eq:num-source-charge}
\end{align}
which gives $Q_{\mathrm{full}}=-8.22$ in our example.  The corresponding massless tangential profile has the asymptotic form
\begin{align}
  \eta(\rho)
  =
  -\frac{Q_\mathrm{full}}{4\pi^2\rho^2}
  +\order{\rho^{-4}}.
  \label{eq:num-massless-tail}
\end{align}
Matching the massive tail to the massless core gives the small-$\mu_\phi^2$ behavior
\begin{align}
  B(\mu_\phi^2)
  =
  B(0)
  +\frac{Q_{\rm full}^2}{32\pi^2}\,
  \mu_\phi^2\ln\left(\frac{1}{\mu_\phi^2}\right)
  +b\mu_\phi^2
  +\dots,
  \label{eq:num-small-delta-general}
\end{align}
where the dots represent higher-order terms in the small $\mu_\phi$ limit. 
Thus $B$ has a finite one-sided limit, while $\dd B/\dd(\mu_\phi^2)$ is logarithmically singular as $\mu_\phi^2\to0^+$.  The corresponding asymptotic fit is shown by the dot-dashed curve in Fig.~\ref{fig:num-massless-limit}, with $b$ determined from the numerical data.

The limit $\mu_\phi^2\to0^+$ removes the positive tangential curvature.  The finiteness of the classical action in this limit does not by itself establish a finite fluctuation prefactor, because the massless tangential mode changes the infrared behavior.  Nevertheless, the result shows that the semiclassical exponential factor remains well defined at the level of the classical bounce action.

We finally comment on a weakly tachyonic tangential curvature, $\mu_\phi^2=-\nu^2<0$.  
In this case, the potential becomes tachyonic in the asymptotic region, causing the long-distance tails to become oscillatory rather than localized. A complete transition probability then requires specifying the initial density matrix and treating the problem in a finite-time Lorentzian or Schwinger--Keldysh formulation.

The massless result can nevertheless provide an estimate of a conditional local bubble-core exponent, provided that the tachyonic evolution is negligible during bubble formation.  Let $R_b$ and $\tau_b$ denote the characteristic bubble radius and formation time.  The tangential mass perturbs the localized core only weakly if $\nu R_b\ll1$ and $\nu\tau_b\ll1$.  At fixed background field value, and provided that no additional approximate zero mode is introduced, we expect
\begin{align}
  B_{\mathrm{core}}(\mu_\phi^2=-\nu^2)
  =
  B_{\mathrm{core}}(0)
  +\order{(\nu R_b)^2,(\nu\tau_b)^2}.
  \label{eq:num-weak-tachyon-core}
\end{align}
In the present example, with lengths measured in units of $m^{-1}$, the profile has $R_b\sim 2$.  Thus $\nu^2=0.01$ gives $(\nu R_b)^2 \sim 0.04$, suggesting a core action close to the massless value, $B_{\rm core}\sim 10.8$.  This estimate does not imply equality of the full finite-time transition probabilities, but it suggests that the local bubble-core exponent remains of the same order for a weakly tachyonic tangential direction.

%%%%%%%%%%%%%%%%%%%%%%%%%%%%%%%%%%%%%%%%%%%%%%%%%%%%%%%%%%%%%%%%%%%%%%%%%%%%%%%%%%%%%%%%%%%%%%%%%%%%
\section{Discussion and conclusions}
\label{sec:discussion}
%%%%%%%%%%%%%%%%%%%%%%%%%%%%%%%%%%%%%%%%%%%%%%%%%%%%%%%%%%%%%%%%%%%%%%%%%%%%%%%%%%%%%%%%%%%%%%%%%%%%

We have formulated an adiabatic tunneling problem for a slowly rolling multi-field background on a fixed on-shell geometry.  
Separating out the background motion of the scalar field renders the effective potential explicitly time dependent. 
For a slow-roll background, this time dependence is parametrically small, and the tunneling problem can be treated by an adiabatic expansion.  
Since the potential is relatively shallow along the slow-roll direction, the tunneling trajectory generically bends toward it, reducing the tunneling action relative to instantaneous transverse tunneling.

For a more rapidly evolving background, tadpole subtraction still provides a useful reformulation of the problem.  The adiabatic expansion, however, is no longer sufficient, and one must instead treat the resulting time-dependent tunneling problem directly~\cite{Widrow:1991xu,Keski-Vakkuri:1996lbi,Draper:2023fkz}.

For a real, time-reflection-symmetric bounce, the correction linear in $\dot{\bphi}_*$ does not contribute to the real tunneling exponent.  The first nonzero correction therefore appears only at second adiabatic order.  This conclusion can fail if the tunneling path is described by genuinely time-dependent collective coordinates.  In that case, the residual one-form discussed in App.~\ref{sec:collective} must be retained.

The gravitational corrections are discussed in App.~\ref{sec:gravity}.  At a stationary background point, the calculation reduces to the standard gravitational vacuum-decay problem, while in the weak-gravity limit it reproduces the flat-space equations.  It does not, however, fully match the scalar momentum and FLRW extrinsic curvature of the physical rolling state.  A complete gravitational treatment is expected to require a complex saddle obtained after reducing the lapse and shift constraints, together with the corresponding phase-space boundary conditions.

\section*{Acknowledgements}
This work was supported by JSPS KAKENHI Grant Number 23K13092.

%%%%%%%%%%%%%%%%%%%%%%%%%%%%%%%%%%%%%%%%%%%%%%%%%%%%%%%%%%%%%%%%%%%%%%%%%%%%%%%%%%%%%%%%%%%%%%%%%%%%
\appendix

\section{Relation to generalized escape paths}
\label{app:GEP}
%%%%%%%%%%%%%%%%%%%%%%%%%%%%%%%%%%%%%%%%%%%%%%%%%%%%%%%%%%%%%%%%%%%%%%%%%%%%%%%%%%%%%%%%%%%%%%%%%%%%

In this appendix we compare tadpole subtraction with the generalized escape-path formulation of Ref.~\cite{Darme:2019ubo}.

\subsection{Generalized escape paths}

We first briefly review the relevant ingredients of the generalized escape-path method.  A semiclassical wavefunctional is written as
\begin{align}
  \Psi[\phi,t]
  =
  \exp\left[
    \frac{\ii}{\hbar}\left(F[\phi,t]+\ii G[\phi,t]\right)
  \right],
  \label{eq:GEP_WKB}
\end{align}
where $F$ controls the phase and $G$ controls the exponential suppression.  Introducing collective coordinates $q^i$ and restricting the configuration space to a surface $\phi_s(\bm x;q^i)$ gives the induced metric
\begin{align}
  g_{ij}(q)
  =
  \int\dd^3x\,
  \frac{\partial\phi_s}{\partial q^i}
  \frac{\partial\phi_s}{\partial q^j}.
  \label{eq:GEP_metric}
\end{align}
For a stationary energy eigenstate, the leading WKB equations restricted to this surface are
\begin{align}
  g^{ij}G_iG_j-g^{ij}F_iF_j&=2(\mathscr U-E),
  \nonumber\\
  g^{ij}F_iG_j&=0,
  \label{eq:GEP_stationary_equations}
\end{align}
where $F_i=\partial_iF$, $G_i=\partial_iG$, $g^{ij}$ is the inverse of $g_{ij}$, and $E$ is the energy eigenvalue.  
Here $\mathscr U(q)$ denotes the potential-energy functional restricted
to the collective-coordinate surface, including the spatial-gradient
energy. 
The $F$ flow describes the classical probability current, while the $G$ flow determines the tunneling suppression.  If both $F_i$ and $G_i$ are nonzero, the orthogonality condition in the second line of Eq.~(\ref{eq:GEP_stationary_equations}) generally cannot be satisfied on a one-dimensional path.  One must instead construct a multidimensional collective-coordinate surface on which the projected saddle condition is satisfied.  This construction is the central practical difficulty in applying generalized escape paths to a moving or time-dependent background.

\subsection{Background phase and residual WKB state}

Tadpole subtraction can be understood as a way of extracting the classical background phase before formulating the residual tunneling problem.  The boundary term in Eq.~(\ref{eq:exact_subtracted_action}) motivates the factorization
\begin{align}
  \Psi[\bphi+\eta,t]
  =
  \exp\left\{
    \frac{\ii}{\hbar}
    \left[
      S_{\mathrm{bg}}^m(t)+\Theta_\phi[\eta;t]
    \right]
  \right\}
  \psi[\eta,t],
  \label{eq:wavefunctional_factorization}
\end{align}
where
\begin{align}
  \Theta_\phi[\eta;t]
  =
  \int\dd^3x\,a^3\dot{\bphi}_I\eta^I.
  \label{eq:scalar_background_phase}
\end{align}
Here $S_{\mathrm{bg}}^m$ is the matter action evaluated on the homogeneous rolling solution $\bphi(t)$, and $\psi$ denotes the residual fluctuation wavefunctional.  Writing
\begin{align}
  \psi[\eta,t]
  =
  \exp\left[
    \frac{\ii}{\hbar}F_{\mathrm{fluc}}[\eta,t]
    -\frac{1}{\hbar}G[\eta,t]
  \right],
  \label{eq:fluctuation_wkb}
\end{align}
the total scalar phase gradient is
\begin{align}
  \frac{\delta F_{\mathrm{tot}}}{\delta\eta^I(\bm x)}
  =
  a^3\dot{\bphi}_I
  +
  \frac{\delta F_{\mathrm{fluc}}}{\delta\eta^I(\bm x)}.
  \label{eq:total_phase_gradient}
\end{align}
The first term is precisely the canonical momentum of the rolling homogeneous background.

In the leading instantaneous treatment, we take the residual fluctuation state to be vacuum-like and set
\begin{align}
  F_{\mathrm{fluc}}^{(0)}=0.
  \label{eq:Ffluc_zero}
\end{align}
After the background phase is extracted, the residual wavefunctional is therefore taken to be real, up to the tunneling suppression, and carries no additional WKB phase.  The functional $G[\eta,t]$ then controls the amplitude such as 
\begin{align}
  |\psi[\eta,t]|^2
  \propto
  \exp\left[-\frac{2}{\hbar}G[\eta,t]\right].
\end{align}
The leading probability exponent for reaching a tunneling configuration $\eta_b$ is
\begin{align}
  B_0(t_*)
  =
  2\Delta G_*
  \equiv
  2\left[
    G_*(\eta_b)-G_*(0)
  \right].
  \label{eq:B_G_relation}
\end{align}
For the real, time-reflection-symmetric frozen bounce considered in Sec.~\ref{sec:frozen}, this quantity is equal to the Euclidean bounce-action difference.

This shows how tadpole subtraction simplifies the generalized escape-path structure in the adiabatic problem.  In the original variables, the total phase contains the rolling background momentum, as in Eq.~(\ref{eq:total_phase_gradient}).  By shifting to $\eta=\phi-\bphi(t)$ and extracting the phase in Eq.~(\ref{eq:scalar_background_phase}), this classical contribution to $F$ is removed from the residual WKB problem.  At leading instantaneous order, where $F_{\mathrm{fluc}}=0$, the orthogonality condition in Eq.~(\ref{eq:GEP_stationary_equations}) is then automatically satisfied.  The price is that the tadpole-subtracted potential, the collective metric, and possible residual one-form terms become explicitly time dependent.

%%%%%%%%%%%%%%%%%%%%%%%%%%%%%%%%%%%%%%%%%%%%%%%%%%%%%%%%%%%%%%%%%%%%%%%%%%%%%%%%%%%%%%%%%%%%%%%%%%%%
\section{Collective coordinates and the residual one-form}
\label{sec:collective}
%%%%%%%%%%%%%%%%%%%%%%%%%%%%%%%%%%%%%%%%%%%%%%%%%%%%%%%%%%%%%%%%%%%%%%%%%%%%%%%%%%%%%%%%%%%%%%%%%%%%

Collective coordinates provide a finite-dimensional description of the field-theory tunneling problem.  In a time-dependent profile basis, however, tadpole subtraction does not remove all velocity-linear terms. A residual one-form remains in the reduced system.

Consider a family of fluctuation profiles parametrized by collective coordinates $q^\alpha(t)$,
\begin{align}
  \eta^I(t,\bm x)
  &=
  \eta^I(\bm x;q^\alpha(t),t),
  \nonumber\\
  e_\alpha^I&\equiv\partial_\alpha\eta^I,
  \qquad
  \eta_t^I\equiv\left.\partial_t\eta^I\right|_q .
  \label{eq:collective_ansatz}
\end{align}
At fixed time, we define the inner product
\begin{align}
  \inner{X}{Y}_t
  =
  \int\dd^3x\,a^3(t)\delta_{IJ}X^IY^J .
  \label{eq:time_inner_product}
\end{align}
The full field velocity decomposes as
\begin{align}
  \dot\phi^I
  =
  \dot{\bphi}^I+\eta_t^I+e_\alpha^I\dot q^\alpha .
  \label{eq:A_velocity_decomposition}
\end{align}
Before extracting the background phase, the coefficient of the term linear in $\dot q^\alpha$ is
\begin{align}
  A_\alpha^{\mathrm{raw}}
  =
  \inner{\dot{\bphi}}{e_\alpha}_t
  +
  \inner{\eta_t}{e_\alpha}_t .
  \label{eq:A_raw_appendix}
\end{align}
The first term is an exact one-form on collective-coordinate space.  Indeed, defining
\begin{align}
  \Lambda(q,t)
  =
  \inner{\dot{\bphi}}{\eta}_t,
  \label{eq:A_Lambda}
\end{align}
we have
\begin{align}
  \partial_\alpha\Lambda
  =
  \inner{\dot{\bphi}}{e_\alpha}_t .
  \label{eq:A_exact_appendix}
\end{align}
The phase transformation generated by $\Lambda$ is the collective-coordinate version of Eq.~(\ref{eq:scalar_background_phase}).  It removes the direct projection of the homogeneous rolling momentum and leaves the residual one-form
\begin{align}
  A_\alpha^{\mathrm{res}}
  =
  \inner{\eta_t}{e_\alpha}_t .
  \label{eq:residual_A}
\end{align}
Thus the rolling background momentum is already absorbed in the tadpole-subtracted representation.  A nonzero one-form remains only if the profile basis itself depends explicitly on time at fixed $q^\alpha$.

Substituting Eq.~(\ref{eq:collective_ansatz}) into the phase-subtracted action gives the effective collective-coordinate Lagrangian
\begin{align}
  L_{\eff}
  =
  \frac12M_{\alpha\beta}(q,t)\dot q^\alpha\dot q^\beta
  +A_\alpha^{\mathrm{res}}(q,t)\dot q^\alpha
  -U_{\eff}(q,t),
  \label{eq:collective_lagrangian}
\end{align}
where
\begin{align}
  M_{\alpha\beta}
  &=
  \inner{e_\alpha}{e_\beta}_t,
  \label{eq:collective_metric}
  \\
  U_{\eff}
  &=
  \mathscr U_{\ts}[\eta;t]
  -\frac12\inner{\eta_t}{\eta_t}_t .
  \label{eq:collective_scalar_potential}
\end{align}
Here $\mathscr U_{\ts}$ includes both the spatial-gradient energy and the tadpole-subtracted potential energy.  The last term in Eq.~(\ref{eq:collective_scalar_potential}) accounts for the explicit motion of the profile basis at fixed collective coordinates.  Therefore the residual one-form in Eq.~(\ref{eq:collective_lagrangian}) must be retained whenever the tunneling ansatz has explicit time dependence beyond that of the homogeneous rolling background.

%%%%%%%%%%%%%%%%%%%%%%%%%%%%%%%%%%%%%%%%%%%%%%%%%%%%%%%%%%%%%%%%%%%%%%%%%%%%%%%%%%%%%%%%%%%%%%%%%%%%
\section{Gravitational backreaction}
\label{sec:gravity}
%%%%%%%%%%%%%%%%%%%%%%%%%%%%%%%%%%%%%%%%%%%%%%%%%%%%%%%%%%%%%%%%%%%%%%%%%%%%%%%%%%%%%%%%%%%%%%%%%%%%

In this appendix we formulate the auxiliary $O(4)$-symmetric gravitational problem obtained by freezing the rolling background.  This construction includes the curvature of the Euclidean bubble and its self-gravity, but it does not solve the full phase-space matching problem for a physical rolling FLRW state.

We define the potential-curvature scale associated with the frozen tadpole-subtracted potential by
\begin{align}
  H_{V*}^2
  =
  \frac{U_{\TS,*}(0)}{3\mpl^2}.
  \label{eq:auxiliary_false_Hubble}
\end{align}
The physical rolling background instead satisfies
\begin{align}
  H_*^2
  &=
  H_{V*}^2+\frac{\delta_{IJ} \dot{\bphi}_{*}^I \dot{\bphi}_*^J}{6\mpl^2},
  \nonumber\\
  \frac{H_*^2-H_{V*}^2}{H_*^2}
  &=
  \frac{\epsilon_{H*}}{3},
  \qquad
  \epsilon_H\equiv -\frac{\dot H}{H^2},
  \label{eq:physical_auxiliary_Hubble_mismatch}
\end{align}
where $\mpl$ is the reduced Planck mass. 
Adding the rolling kinetic energy to $U_{\TS,*}(0)$ would not remove this mismatch, because kinetic energy has non-vacuum pressure and cannot be represented by a constant Euclidean potential.

We first state sufficient conditions under which the local flat-space equations used in the main text are valid.  These conditions require both the background curvature and the bubble self-gravity to be weak.  We define the characteristic stress scale inside the bubble by
\begin{align}
  \Delta\mathcal T_{*}
  =
  \max_{\mathrm{bubble}}
  \left\{
    |\Delta U_{\TS,*}(\eta)|,
    \frac12|\partial\eta|^2
  \right\}.
  \label{eq:characteristic_stress_excursion}
\end{align}
A sufficient decoupling regime for a bubble of radius $R_b$ is
\begin{align}
  |H_{V*}^2|R_b^2\ll1,
  \qquad
  \frac{\Delta\mathcal T_{*}R_b^2}{\mpl^2}\ll1.
  \label{eq:weak_gravity_conditions}
\end{align}
For a thin-wall bubble with wall tension $S_1$, one should also require $S_1R_b/\mpl^2\ll1$.

When these conditions are not satisfied, gravitational backreaction must be included.  The auxiliary Euclidean action is
\begin{align}
  S_{E,*}^{\mathrm{aux}}
  ={}&\int_{\mathcal M}\dd^4x\sqrt g
  \left[
    -\frac{\mpl^2}{2}R
    +\frac12\delta_{IJ}\partial_\mu\eta^I\partial^\mu\eta^J
    +U_{\TS,*}(\eta)
  \right]
  \nonumber\\
  &+ S_{\rm GHY} ,
  \label{eq:auxiliary_Euclidean_action}
\end{align}
where $S_{\rm GHY}$ is the Gibbons-Hawking-York boundary term.  
For an \texorpdfstring{$O(4)$}{O(4)}-symmetric ansatz,
\begin{align}
  \dd s_E^2
  =
  \dd\xi^2+\varrho(\xi)^2\dd\Omega_3^2,
  \label{eq:O4_gravitational_metric}
\end{align}
the reduced action becomes
\begin{align}
  S_{E,*}^{\mathrm{aux}}
  =
  2\pi^2\int\dd\xi
  \left\{
    \varrho^3
    \left[
      \frac12\eta_I'\eta^{I\prime}
      +U_{\TS,*}(\eta)
    \right]
    -3\mpl^2\varrho(\varrho'^2+1)
  \right\}.
  \label{eq:reduced_gravitational_action}
\end{align}
The scalar equations, the Hamiltonian constraint, and the scale-factor equation are
\begin{align}
  \eta^{I\prime\prime}
  +3\frac{\varrho'}{\varrho}\eta^{I\prime}
  &=
  U_{\TS,*}^{,I},
  \label{eq:CDL_scalar_equation}
  \\
  \varrho'^2
  &=
  1+\frac{\varrho^2}{3\mpl^2}
  \left(
    \frac12\eta_I'\eta^{I\prime}
    -U_{\TS,*}(\eta)
  \right),
  \label{eq:CDL_constraint}
  \\
  \varrho''
  &=
  -\frac{\varrho}{3\mpl^2}
  \left(
    \eta_I'\eta^{I\prime}
    +U_{\TS,*}(\eta)
  \right).
  \label{eq:CDL_scale_factor_equation}
\end{align}
For a compact de Sitter-type solution, regularity at the two poles requires
\begin{align}
  &\varrho(0)=0,
  \quad
  \varrho'(0)=1,
  \quad
  \eta^{I\prime}(0)=0,
  \nonumber\\
  &\varrho(\xi_{\max})=0,
  \quad
  \varrho'(\xi_{\max})=-1,
  \quad
  \eta^{I\prime}(\xi_{\max})=0.
  \label{eq:compact_CDL_boundary_conditions}
\end{align}
For $U_{\TS,*}(0)>0$, the auxiliary false-vacuum solution is
\begin{align}
  \eta^I=0,
  \qquad
  \varrho_F(\xi)=H_{V*}^{-1}\sin(H_{V*}\xi),
  \qquad
  0\leq\xi\leq\frac{\pi}{H_{V*}}.
  \label{eq:auxiliary_false_dS_solution}
\end{align}
The corresponding auxiliary gravitational exponent is
\begin{align}
  B_{\mathrm{grav}}(t_*)
  =
  S_{E,*}^{\mathrm{aux}}[\eta_b,\varrho_b]
  -
  S_{E,*}^{\mathrm{aux}}[0,\varrho_F],
  \label{eq:auxiliary_gravitational_exponent}
\end{align}
where $(\eta_b,\varrho_b)$ is the bounce solution.  Using the constraint in Eq.~(\ref{eq:CDL_constraint}), and then the scale-factor equation together with
the compact boundary conditions, 
the on-shell action can be written as
\begin{align}
  S_{E,*}^{\mathrm{aux}}[\eta_b,\varrho_b]
  &=
  4\pi^2\int\dd\xi
  \left[
    \varrho_b^3 U_{\TS,*}(\eta_b)
    -3\mpl^2\varrho_b
  \right]
  \nonumber\\
  &=
  -2\pi^2\int\dd\xi\,
  \varrho_b^3 U_{\TS,*}(\eta_b).
  \label{eq:gravitational_onshell_action}
\end{align}
For the auxiliary false-vacuum solution,
\begin{align}
  S_{E,*}^{\mathrm{aux}}[0,\varrho_F]
  =
  -\frac{24\pi^2\mpl^4}{U_{\TS,*}(0)}.
  \label{eq:false_deSitter_action}
\end{align}

There are two useful checks of this construction.  First, for $\mpl\to\infty$ at fixed scalar scales, one has $\varrho(\xi)=\xi+\order{\mpl^{-2}}$, and Eq.~(\ref{eq:auxiliary_gravitational_exponent}) reduces to the flat-space exponent in Eq.~(\ref{eq:O4_action}).  Second, if $V_{,I}(\bphi_*)=0$, then $U_{\TS,*}=V$ up to an irrelevant constant, and the auxiliary problem becomes the ordinary Coleman--De~Luccia problem~\cite{Coleman:1980aw}.

The auxiliary solution above should not be identified with the full gravitational instanton for the rolling system.  In the original variables, $V_{,I}(\bphi_*)\neq0$, and hence the constant configuration $\phi^I=\bphi_*^I$ is not a Euclidean solution.  The full decay problem must match the scalar momentum represented by Eq.~(\ref{eq:scalar_background_phase}) together with the FLRW extrinsic curvature of the rolling background.  The corresponding saddle is therefore generally expected to be complex and need not be $O(4)$ symmetric.  Its action should be compared with the physical rolling FLRW history on the same complex contour, rather than with the auxiliary false-vacuum action in Eq.~(\ref{eq:false_deSitter_action}).

\bibliography{reference}

\end{document}